\documentclass[useAMS,fleqn,usenatbib]{mnras}
\usepackage{newtxtext,newtxmath}
\usepackage{multirow}
\usepackage[T1]{fontenc}
\usepackage{ae,aecompl}
\usepackage{todonotes}
\usepackage{graphicx}	
\usepackage{amsmath}	
\usepackage{amssymb}	
\usepackage{threeparttable}
\usepackage{booktabs}
\newcommand{\ra}[1]{\renewcommand{\arraystretch}{#1}}
\def\apj {ApJ}
\def\apjl {ApJL}
\def\apjs {ApJS}
\def\aj {AJ}
\def\aap {A\&A}
\def\mnras {MNRAS}
\def\araa {ARA\&A}
\def\nat {Nature}

\def\Msol {M_{\odot}}

\def\R200 {R_{200}}
\def\Lx {L_{X}}
\def\mpc{h^{-1} {\rm{Mpc}}}
\def\kms {\rm{km~s^{-1}}}
\def\Mpc {\rm Mpc}

\def\Gyr {\rm Gyr}

\def\ergs {\rm erg~s^{-1}}

\title[Anisotropic infall in clusters]{Anisotropic infall in the outskirts of OmegaWINGS 
galaxy clusters}

\author[Salerno et al.]{\parbox[t]{\textwidth}{
Juan Manuel Salerno$^{1}$\thanks{jsalerno@oac.unc.edu.ar},
H\'ector J. Mart\'inez$^{1,2}$, Hern\'an Muriel$^{1,2}$, Valeria Coenda$^{1,2}$,
 Benedetta Vulcani$^{3}$, Bianca Poggianti$^{3}$, Alessia Moretti$^{3}$, Marco Gullieuszik$^{3}$, Jacopo Fritz$^{4}$, and Daniela Bettoni$^{3}$}
 \\
 \\
$^{1}$Instituto de Astronom\'{\i}a Te\'orica y Experimental (IATE), CONICET - UNC, Laprida 854, X5000BGR, C\'ordoba, Argentina\\
$^{2}$Observatorio Astron\'omico, Universidad Nacional de C\'ordoba, Laprida 854, X5000BGR, C\'ordoba, Argentina\\
$^{3}$INAF - Osservatorio Astronomico di Padova, Vicolo Osservatorio 5, IT-35122 Padova, Italy\\
$^{4}$Instituto de Radioastronom\'ia y Astrof\'isica, IRyA, UNAM, Campus Morelia, A.P. 3-72, C.P. 58089, Mexico
}

\date{Accepted 2020 February 16. Received 2020 2020 January 29; in original form 2019 November 15}

\pubyear{2019}

\hypersetup{draft} 
\begin{document}
\label{firstpage}
\pagerange{\pageref{firstpage}--\pageref{lastpage}}
\maketitle
 
\begin{abstract}
We study the effects of the environment on galaxy quenching in the outskirts of clusters at 
$0.04 < z < 0.08$. We use a subsample of 14 WINGS and OmegaWINGS clusters that are linked
to other groups/clusters by filaments and study separately galaxies located in two regions in the 
outskirts of these clusters according to whether they are located towards the filaments' directions
or not. We also use samples of galaxies in clusters and field as comparison. 
Filamentary structures linking galaxy groups/clusters were identified over the 
Six Degree Field Galaxy Redshift Survey Data Release 3. 
We find a fraction of passive galaxies in the outskirts of clusters intermediate between that
of the clusters and the field's. We find evidence of a more effective quenching in the
direction of the filaments. 
We also analyse the abundance of post-starburst galaxies in the outskirts of clusters
focusing our study on two extreme sets of galaxies according to their phase-space position:
backsplash and true infallers. 
We find that up to $\sim70\%$ of post-starburst galaxies in the direction of filaments are likely 
backsplash, while this number drops to $\sim40\%$ in the isotropic infall region.
The presence of this small fraction of galaxies in filaments that are falling into clusters for the first time 
and have been recently quenched, supports a scenario in which a significant
number of filament galaxies have been quenched long time ago.
\end{abstract}


\section{Introduction}
\label{sect:Intro}
It is well known that the environment affects the properties of galaxies such as star 
formation, morphology, luminosity, colour, gas content, and the structure
of their subsystems. In dense environments, galaxies tend to be massive, 
red, early-type, with little or no on-going star formation activity, inhabiting the 
red sequence of the diagram color-magnitude (CMD). On the other hand, a population of 
less massive, blue, late-type, and star forming galaxies that lie on the blue cloud 
of the CMD, is found in low density environments. Galaxy morphology 
correlates strongly with the local galaxy density or the distance from the cluster
centre (e.g. \citealt{dressler80}, \citealt{Whitmore:1993}, \citealt{Dominguez:2001}, \citealt{Bamford:2009}, 
\citealt{Paulino-Afonso:2019}). Colour and luminosity have been proposed as two of 
the properties that predict best the environment (e.g. \citealt{Blanton05}, 
\citealt{martinez06}, \citealt{martinez08}). Moreover, the CMD correlates with the 
environment (e.g.  \citealt{Schawinski:2007a}, \citealt{Martinez:2010}, 
\citealt{Cooper:2010b}, \citealt{coenda18}) and morphological types (e.g. 
\citealt{Takamiya:1995}, \citealt{Schawinski:2009}, \citealt{Masters:2010}). This is 
related to the dependence of the fraction of star-forming galaxies on the environment 
(e.g. \citealt{Hashimoto}, \citealt{Mateus:2004}, \citealt{Rines:2005}, 
\citealt{BlantonMoustakas:2009}, \citealt{Welikala:2008}, \citealt{Schaefer:2017}, 
\citealt{Coenda:2019}). In additon, galaxies in low density environments appear on 
average younger than their counterparts in high-density environments (e.g. 
\citealt{Thomas:2005}, \citealt{Cooper:2010a}, \citealt{Zheng:2017}).

Clusters of galaxies are the most massive objects in virial equilibrium that can be
found in the Universe. These systems grow by the accretion of galaxies and groups of 
galaxies from their outskirts preferentially alongside filaments, and, to a lesser 
extent, from  other directions. The cluster environment is characterised by a deep 
gravitational potential well, and by an intra-cluster medium (ICM) which is filled 
with hot ionised gas. Inside clusters, galaxies have properties that differ from those
in the field. Galaxies passing through the ICM at high velocities suffer ram pressure
stripping that removes an important fraction of their cold gas (\citealt{GG:1972}, 
\citealt{Abadi:1999}), with the consequent decrease of their SFR. 
The trip of a galaxy through the ICM can also
remove the galaxy's warm gas, mechanism known as starvation 
(\citealt{Larson:1980,McCarthy:2008,Bekki:2009,Bahe:2013,Vijayaraghavan:2015}). Starvation 
cuts off the supply of gas that cools out from 
the galaxy's halo, thus inhibiting subsequent star formation. Other mechanisms that 
galaxies experience in their passage through the inner region of clusters which are
responsible of affecting their evolution are: 
tidal stripping (e.g. \citealt{Zwicky:1951,Gnedin:2003a,Villalobos:2014}); thermal 
evaporation (e.g. \citealt{Cowie:1977}); 
and galaxy - galaxy interactions or harassment (e.g. 
\citealt{Moore:1996,Moore:1999,Gnedin:2003b}).

Most of the processes described in the previous paragraph tend to decrease or completely suppress the star formation in galaxies and are generally referred to as galaxy quenching. Besides all the aforementioned environmental sources of quenching, the shut down of the star
formation can also be produced by intrinsic properties of galaxies. These processes are strongly 
correlated with the stellar mass, therefore they are collectively refered to as mass quenching.  
Among the suggested processes are: supernova-driven winds 
(e.g., \citealt{Stringer:2012, Bower:2012}),
 halo heating \citep{Marasco:2012}, and AGN feedback (e.g., \citealt{Nandra:2007, 
Hasinger:2008, Silverman:2008, Cimatti:2013}). 

In the framework of the hierarchical formation of structures, clusters of galaxies are continuously accreting galaxies.  It has been suggested that in this process of falling, galaxies could undergo different physical processes that could affect the star formation even before they reach the cluster. Consequently, to fully understand what the cluster environment produces in galaxies, it is of key
importance to have a throughout characterisation of the population of galaxies in the
outskirts of clusters.
Several observations have shown that properties of galaxies such as star formation, gas
content and colour, are affected by the cluster environment at large clustercentric 
distances (e.g. \citealt{Solanes:2002,Lewis:2002,gomez03,Braglia:2009,Park:2009,Hansen:2009,
vonderLinden:2010,Haines:2015,Rhee:2017}). 
In particular, spiral galaxies with low star formation rates were found in the outskirts of 
clusters in early studies such as \citet{Couch:1998}, or \citet{Dressler99}. In recent years,
a deficit of star forming galaxies in the infalling region of clusters has been reported 
(e.g. \citealt{Wetzel:2013,Haines:2015,Bianconi:2018}). This has been reproduced in
simulations by \citet{Bahe:2013}. These results can be explained by the presence of 
environmental effects accelerating the consumption of the gas reservoir before galaxies 
enter in a cluster, a process known as pre-processing (e.g. 
\citealt{Mihos:2004,Fujita:2004}). 
An important fraction of the cluster galaxies have spent time in 
groups or filaments before they fall into the cluster (e.g. \citealt{McGee:2009,deLucia:2012,Wetzel:2013,Hou:2014}). The population of galaxies
in the outskirts of clusters includes not only galaxies that have not yet entered the
cluster but also backsplash galaxies, i.e., galaxies that have passed close to the centre 
of the cluster since their infall and are now beyond the virial radius 
(e.g. \citealt{Mamon:2004,Gill:2005,Mahajan:2011}).
For an adequate characterisation of the properties of galaxies that are 
falling into clusters, it is important to take into account the contamination by 
backsplash galaxies, which, having orbited through the inner regions of a cluster, 
could have been affected by the physical processes present in that extreme 
environment. The backsplash scenario 
in the evolution of galaxies has also been explored in \citet{Rines:2005}, 
\citet{Pimbblet:2006}, \citet{Aguerri:2010} and \citet{Muriel:2014}.

Some studies have analysed the infall region considering all possible directions,
while other works consider the fall through filaments only. The ESO Distant Cluster Survey 
\citep{White:2005} has been used to study the infall region of clusters in the redshift 
range $0.4 < z < 0.8$ \citep{just15}. They find that the fraction of red galaxies in the 
outskirts of clusters is intermediate between the field and cluster values, and conclude 
that pre-processing may have already started during the fall. In 
addition, \citet{Bianconi:2018} analyse a galaxy cluster sample down from the Local 
Cluster Substructure Survey, and find that the fraction of star forming galaxies in 
infalling groups, is lower as a function of the clustercentric distance than in the overall
galaxy population in clusters. They consider this as an evidence of pre-processing of
galaxies within groups that are falling into galaxy clusters. \citet{Einasto:2018} have studied the
structure and galaxy population in the cluster A2142 and its outskirts. They propose that this cluster 
has been formed as a result of infalling groups and past and recent mergers, along the filament axis. 

Using cosmological simulations, \citet{Zinger:2018} have found that the hot gas is removed 
from galaxy halos, and this process is more effective between 1 and 3 virial radii. 
They argue that the removal of gas from the galaxy's halo sets the stage for the quenching of the star formation by starvation over 2-3 Gyr before it enters the cluster. Although it is a slow process (the gas in the disc cannot be replenished and the star formation in the galaxy will eventually cease), it anticipates the quenching that will be completed once  the galaxy enters the cluster.

There are several works that focus the analysis on the properties of galaxies in the filament
region. \citet{Kraljic18} have studied the impact of the large-scale environment on galaxy properties 
in the nearby Universe, using the Galaxy And Mass Assembly (GAMA, \citealt{Driver:2009}). 
They identify peaks of density, filaments,
and walls, in the distribution of galaxies. They find that the fraction of red galaxies
increases when approaching to the peaks of density within filaments. 
Using the same survey, \citet{Alpaslan:2016} have found that isolated spiral galaxies 
have higher stellar masses and lower SFR towards the inner regions of filaments than in 
the filament periphery. 
Similarly, \citet{Laigle17} study the properties of galaxies with photometric redshifts
in the range $0.5 < z_{\rm phot} < 0.9$ and find that, at fixed stellar mass, 
passive galaxies are more confined towards the core of filaments. 

Using the Sloan Digital Sky Survey (SDSS, \citealt{York:2000}), \citet{Chen17} find that at 
the same local density, galaxies that reside closer to filaments are more massive than those 
further out, in agreement with \citet{Alpaslan:2016} and \citet{Laigle17}. 
Using the same survey, but at intermediate redshifts, $0.12 < z < 0.40$, \cite{zhang13} 
detect an evolution in the blue fraction of filament galaxies located in between pairs of 
clusters, that is not observed in clusters. At higher redshift ($z \sim 0.8$), 
\citet{Malavasi17} use the VIPERS survey \citep{Guzzo14} and find a segregation in star formation, in 
the sense that star-forming galaxies are preferentially located in the outskirts of 
filaments.  A different result was reported by \citet{Darvish14}, who find at $z = 0.845$ an 
intrinsic enhancement of the fraction of $H\alpha$ emitters in filaments with respect to clusters and the field.
More recently, \cite{Vulcani19} analyse the spatial distribution of $H\alpha$ emission of galaxies in filaments and find four galaxies with recent star formation. 
They hypothesize  that  these  galaxies are passing  through filaments that are able to cool the gas and increase the star formation.

Galaxies are accreted to clusters mainly through filaments (e.g. \citealt{Colberg:1999,Ebeling:2004}), and to a lesser extent from other directions.
A question that arises is whether, depending on the infall direction, they experience 
different processes. In other words, whether the properties of galaxies in the outskirts of 
clusters are the same in all directions, or, on the contrary, they depend on the presence 
of filaments linking clusters. \citet{Martinez16} select filaments between galaxy groups in the SDSS up to $z = 0.15$.  They distinguish whether galaxies are falling into groups along filaments or from other directions (isotropic infalling). They find that 
filaments play a specific role in quenching galax ies that is revealed when comparing
to the isotropic infall region. A similar analysis has been recently done using VIPERS 
data in the redshift range $0.43 < z < 0.89$ by \citet{Salerno2019}, finding similar 
results to the low redshift analysis by \citet{Martinez16}, and showing that the filament 
environment was able to effectively quench galaxies as early as $z \sim 0.9$. 

Our goal is to study the effects of environment upon galaxies that are falling into rich clusters.
We distinguish between galaxies that are infalling in the directions defined by filaments from 
those that are being accreted isotropically. Studying the peripheries of galaxy clusters 
has a serious drawback in the contamination by interloper galaxies, making necessary the use of spectroscopic redshifts. The OmegaWINGS spectroscopic survey (\citealt{Gullieuszik15}; \citealt{Moretti17} ) enlarges the 
number of cluster members of WINGS clusters (\citealt{Fasano06}; \citealt{Moretti14}) out to large radii, reaching the virial radius and beyond. 
Therefore, these surveys constitute an excellent sample for studying the outskirts of
galaxy clusters.

This article is organised as follows: we describe the galaxy data in Sect. \ref{sect:data}; Sect. 
\ref{sect:filam} deals with the identification of filaments and the resulting cluster sample; the 
outskirts of clusters are described in Sect. \ref{sect:out}. In Sect. \ref{sect:results} we compare 
the fraction of passive (\ref{sect:emlpas}), emission line (\ref{sect:emlpas}), and post-starburst 
(\ref{sect:ps}) galaxies, in clusters, field, filaments, and the isotropic infall region. 
We also discuss the implications of our results in that section. Finally, we summarise our main 
results in Sect. \ref{sect:conclu}. Throughout the paper we assume a flat cosmology with density 
parameters $\Omega_{\rm m} = 0.30$, $\Omega_{\Lambda} = 0.70$, and a Hubble's constant 
$H_0 = 70 \, \kms$ $\Mpc^{-1}$.

\section{Samples}
\subsection{WINGS and OmegaWINGS clusters and galaxies}
\label{sect:data}
Clusters of galaxies used in this paper are a subsample drawn from the OmegaWINGS spectroscopic survey 
(hereafter OW, \cite{Gullieuszik15,Moretti17}. This survey is an extension of the WIde-field Nearby 
Galaxy-cluster Survey (WINGS, \citealt{Fasano06,Moretti14}), a multi-wavelength survey designed to cover the 
outskirts of 76 rich clusters, with redshift range $0.04\leq z \leq 0.07$ that were selected from the 
ROSAT All Sky Survey (\citealt{Ebeling96}).

OmegaWINGS extends the WINGS survey in terms of cluster spatial coverage for 46 of these clusters, imaging in the U, B, and V bands have been obtained covering an area 
of $\approx$ 1 deg$^2$.  Also, it covers a wide range of velocity dispersion $\sigma \sim 500-1300 
\,\kms$,  X-ray luminosity $\Lx \sim 0.2-5 \times 10^{44}\, \ergs$,  and $\R200 \sim 1-3 \, \Mpc$.
The target selection was similar for the two surveys \citep{Cava09,Moretti17}.  They were selected to 
have a total magnitude brighter than $V = 20$, excluding only those well above the color-magnitude 
sequence with $B- V > 1.20$.
We selected the galaxies brighter than $M_V=-17.4$ (the absolute magnitude limit of a galaxy with V=20 in the more distant cluster of the sample, see \citet{Paccagnella17}).  

The spectroscopic redshifts were computed  with a semiautomatic method, which involves the automatic 
cross correlation technique and the emission lines identification, with a very high success rate 
($\approx 95\%$ for the whole sample, see \citealt{Cava09} and \citealt{Moretti17}).
To establish the correction for incompleteness (geometrical and magnitude) in the  spectroscopic 
catalogues,  they use the ratio of number of spectra yielding a redshift to the total number 
of galaxies 
in the parent photometric catalogue, calculated both as a function of V magnitude and radial projected 
distance from the brightest cluster galaxy (BCG).
All our calculations have been done weighing each galaxy to correct for both incompletenesses.

The final spectroscopic sample includes 14,801 galaxies in the 35 out of 46 OmegaWINGS clusters that have
spectroscopic completeness higher than $50\%$.
Galaxy properties (stellar masses and spectral
types) have been derived by fitting the fiber spectra with SINOPSIS (SImulatiNg OPtical Spectra wIth
Stellar population models), a spectrophotometric model fully described in \cite{Fritz07,Fritz2014}. Stellar masses are computed using the initial mass function proposed by  \cite{Salpeter55}.  

The OW galaxies have been classified according to their spectral characteristics in three different types: 
passive (PAS), emission-line (EML) and post-starburst (PS). Those galaxies classified as PS are likely to have their 
star formation suddenly truncated at some point during the last $0.5-1.0 \,\Gyr$. 
For more details on the spectral classification we refer the reader to \cite{Fritz2014} and  \cite{Paccagnella17}.

\subsection{Filaments of galaxies}
\label{sect:filam}

The goal of this paper is to understand the effects of the filaments upon galaxies by comparing 
the properties of OW galaxies in the outskirts of clusters taking into account whether they are 
in filaments, or they are infalling from other directions. Therefore, we 
select a subsample of OW clusters for which we are able to determine they are nodes of 
filamentary structures.

 \begin{figure}
\includegraphics[width=\hsize]{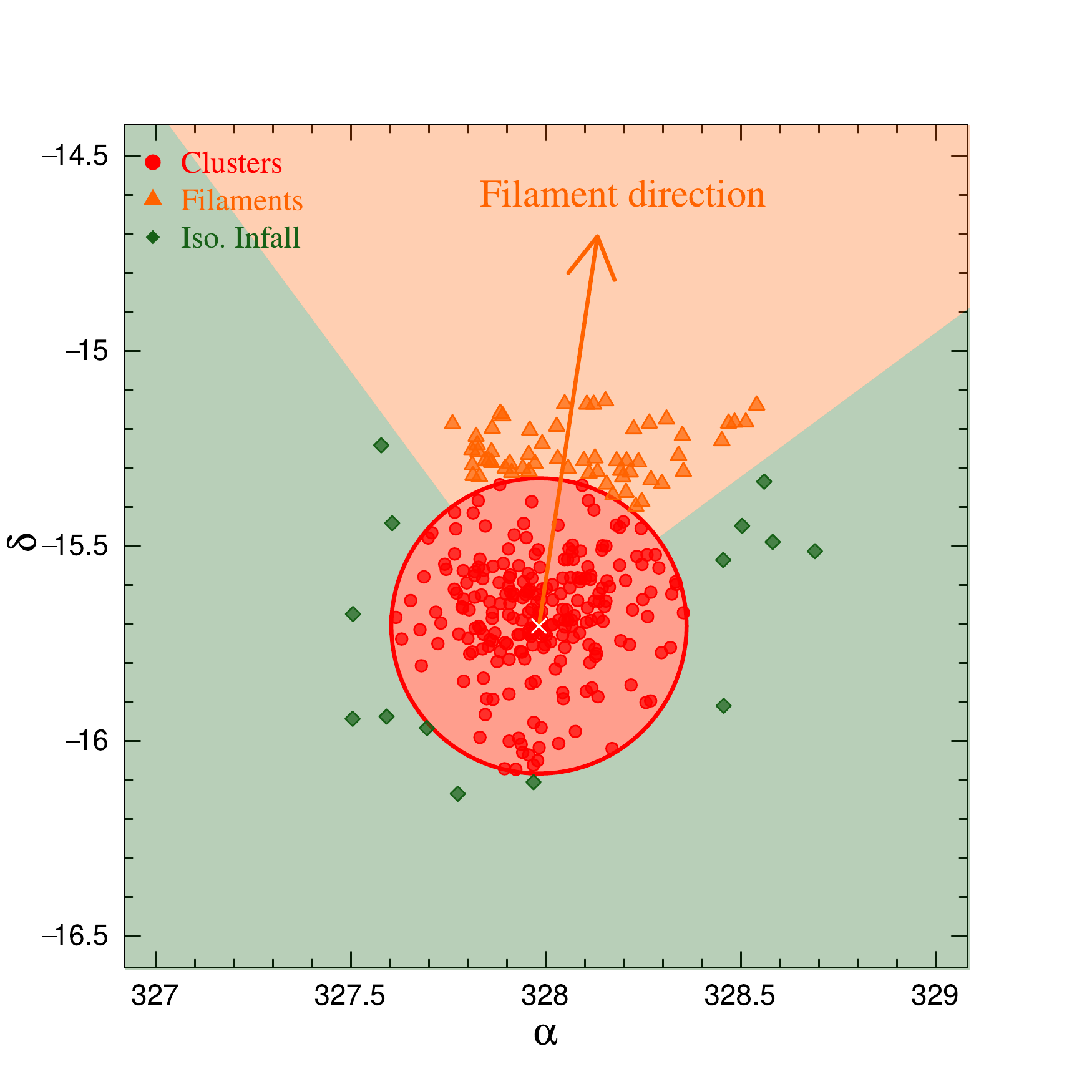}
\caption{An example of the different environments in an OmegaWINGS cluster. 
The circle corresponds to $ \R200 $. The orange arrow indicates the direction towards 
the filament. Galaxies in the cluster are shown as filled red circles. 
Isotropically infalling galaxies are shown as green diamonds, while orange triangles 
represent galaxies in the filament region.}
\label{fig:regions}
\end{figure}
\begin{figure*}
\begin{center}
\includegraphics[width=13 cm]{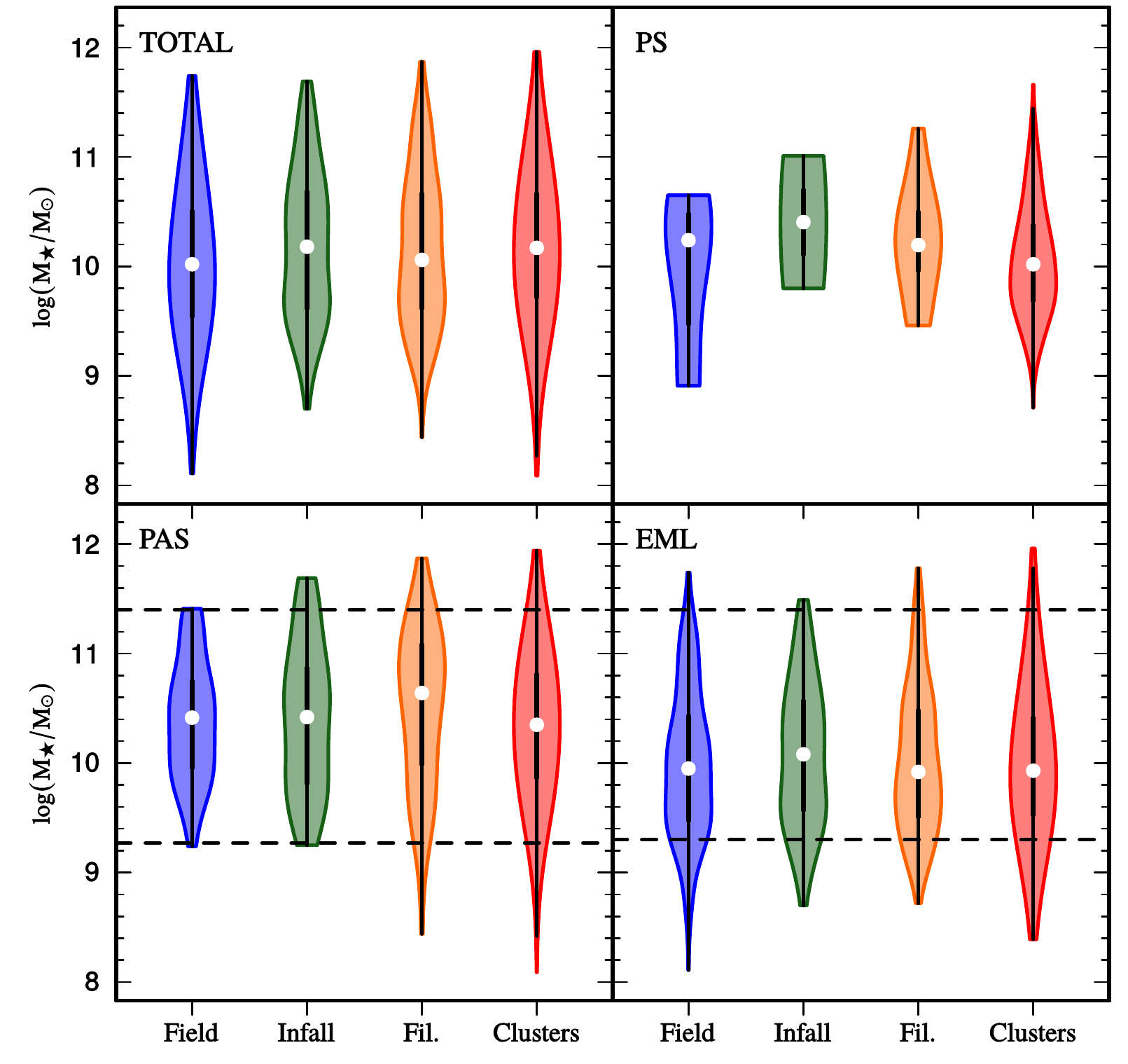}
\caption{ Violin plots of the weighted mass distributions of our samples of galaxies. The box plot 
inside each violin shows the interquartile range. Inner dots in the box plots represent the 
medians of the distributions.
The widths of the violin plots are scaled by the number of galaxies in each bin. \emph{Top left} panel corresponds to the total sample and \emph{top right} panel 
shows post-starburt galaxies.
\emph{Bottom left} considers passive galaxies while 
\emph{bottom right} panel emission-line galaxies. The dashed lines in the bottom panels correspond to the low and high mass limits of $\log(M_{\star}/M_{\odot})=9.3$ and  $\log(M_{\star}/M_{\odot})=$11.4 respectively (see the text). Colours detach environments: blue for field, green for isotropic infall, orange for filaments, and red for clusters of galaxies.}
\label{fig:mass}
\end{center}
\end{figure*}

Filament identification was performed in an independent sample of groups/clusters (hereafter systems) in the area surveyed by the Six Degree Field Galaxy 
Survey\footnote{http://www-wfau.roe.ac.uk/6dFGS/index.html}
(6dFRS, \citealt{Jones:2004}) Data 
Release 3 (DR3, \citealt{Jones09}) following the procedure of 
\citet{Martinez16}\footnote{A paper on this filament sample is currently in preparation
(Mart\'inez et al. 2020).}:
\begin{enumerate}
    \item We use as potential nodes of filaments those systems identified by \citet{Lim17} 
    over the 6dFGRS DR3, that are more massive than $10^{13}\Msol $.
    By choosing this low-mass cut-off we are selecting all massive systems 
    while reducing the contamination by spurious systems.
    \item We search for all pairs of systems that are separated by less than $14~\Mpc$ in redshift space. This is equivalent to $10~\mpc$ in \citet{Martinez16} with $h=0.7$, which was chosen to be similar to the characteristic length of the redshift space two-point correlation function of groups this massive \citep{Zandivarez:2003}. These pairs are candidates to be linked by filaments.
    \item We compute the projected galaxy overdensity in cuboid-like regions defined in redshift 
    space in the region between the candidate nodes (see \citealt{Martinez16} for details). We consider 
    a pair of candidate nodes to be linked by a filament if the overdensity is $(n_T-n_R)/n_R>1$, where
    $n_T$ and $n_R$ are the number of tracer and (normalised) random galaxies in the region, respectively.
\end{enumerate}

To compute galaxy overdensities in the item (iii) above, we use a sample of tracer 
galaxies that includes all 6dFGRS  galaxies brighter than $K_s=12.65$ that have redshift 
quality equal to 3 or 4 (see \citealt{Jones09}). This constitutes a 
reliable sample of galaxies with well established completness criteria. 
We construct a random catalogue based on these tracer galaxies (100 times denser) following 
the cloning procedure developed by \citet{Cole:2011}. For this purpose, we constructed an 
angular coverage mask using the original targets selected for redshift measurement in 6dFGRS, the tracer sample, and routines from the software 
HEALPix\footnote{http://healpix.sourceforge.net} package \citep{Gorski:2005}.

We adopt this overall approach to identify which OmegaWINGS clusters are nodes of filaments since it has proven useful in studying anisotropic infall around
systems \citep{Martinez16,Salerno2019}. We are interested in probing the 
outskirts of clusters
and not in the study of filaments at larger scales, or their characteristics such as 
shape. The sample of systems of \citet{Lim17} is one of the most complete covering the 6dFGRS, 
this enhances the chances of identifying filaments that link systems in the survey.

The sample of filaments comprises 808 filaments in the area covered by 6dFGRS, and in 
the redshift range $0<z<0.1$.
It is important to note that the 6dFRS was designed as a redshift survey 
with the main objective of studying the large-scale structure, and therefore it is 
ideal to identify the filaments that connect systems of galaxies. 
The OmegaWINGS survey, on the other hand, allows for a detailed characterisation of 
the star formation history of galaxies in a 
statistically complete sample of X-ray luminous clusters. We cross-matched OW 
clusters with the sample of filament nodes and 
found that 14 OW clusters are nodes of filaments. These 14 clusters constitute the 
subsample of OW clusters used in this paper.

\begin{table*}\centering
\ra{1.3}
\begin{threeparttable}
\begin{tabular}{lcccccccccccccc}\toprule
           & \multicolumn{2}{c}{TOTAL}  && \multicolumn{3}{c}{PAS}  && \multicolumn{3}{c}{EML} && \multicolumn{3}{c}{PS}\\
            \cmidrule{2-3} \cmidrule{5-7} \cmidrule{9-11} \cmidrule{13-15}
            & $\#$ & $w\#$ && $\#$ & $w\#$ & $wf$       && $\#$ & $w\#$& $wf$   &&$\#$ & $w\#$ & $wf$                   \\ \midrule
Field       & 371 & 628 && 60 & 101  & $0.16\pm0.02$ && 306 & 520& $0.82\pm0.02$ && 5  & 6   & 0.01\tnote{\textdagger}\\
Iso. Infall & 107& 143 && 41  & 52  & $0.52\pm 0.02$ && 68  & 84 & $0.44\pm0.01$   && 5  & 6   & 0.04\tnote{\textdagger}\\
Filaments   & 258 & 318 && 174 & 215 & $0.55\pm0.02$   && 71  & 86 & $0.39\pm0.01$   && 13 & 19  & $0.06\pm0.01$         \\
Clusters    & 4604&6491 && 2572& 3620& $0.58\pm0.01$ && 1683& 2343& $0.32\pm0.01$ && 355& 533  & $0.09\pm0.01$         \\
\bottomrule
\end{tabular}
\begin{tablenotes}
\item[\textdagger]{Due to the low number of galaxies, the boostrap technique could not be applied.}
\end{tablenotes}
\end{threeparttable}
\caption{Number of galaxies ($\#$), weighted number of galaxies ($w\#$) and  weighted
fractions ($wf$), as a function of the environment and spectral type. Errors were computed using the bootstrap resampling technique.}
\label{tab:sample}
\end{table*}

\begin{figure*}
\begin{center}
\includegraphics[width=15 cm]{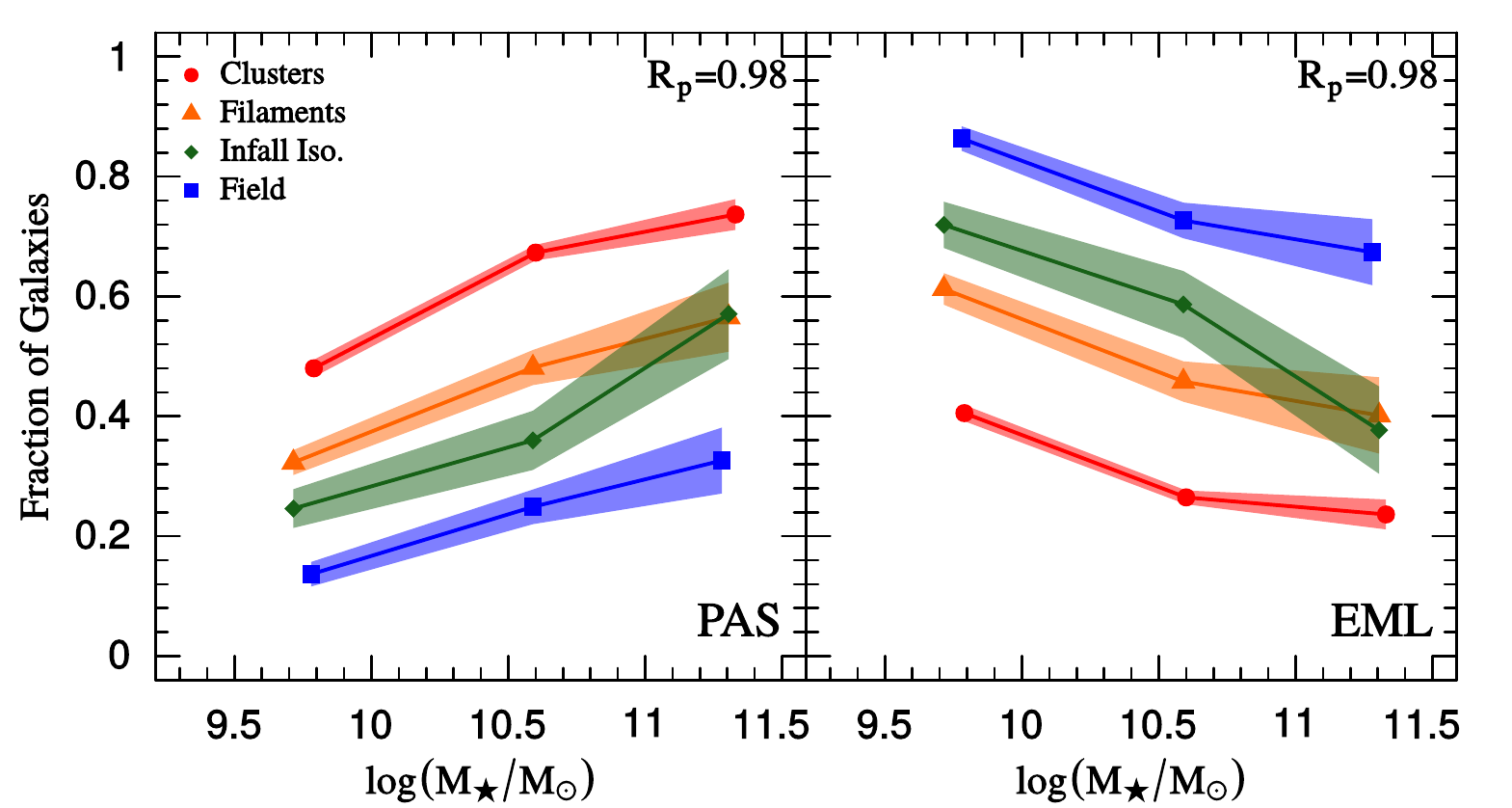}
\caption{ PAS and EML galaxies fraction as a function of 
stellar mass for our samples of galaxies. Solid blue squares correspond to galaxies in the field, green 
diamonds to galaxies falling through isotropic infall, while orange triangles correspond to galaxies 
falling through filaments. Galaxies in clusters are shown in red circles. {\em Left panel} corresponds PAS 
galaxies, and {\em right panel} to the EML galaxies. Error-bars were computed with the bootstrap 
resampling technique. Each panel shows the rejection probability $R_p$ of the null hypothesis (see the text 
for more details).}
\label{fig:frac_mass}
\end{center}
\end{figure*}

\subsection{Galaxies in the outskirts of clusters}
\label{sect:out}

Since the outskirts of clusters are the main objective of this study,
the assignment of galaxies to this environment was performed prioritizing completeness
at the cost of losing purity (i.e. some contamination by field, and cluster galaxies 
is expected). Accordingly, we consider as the outskirt regions of clusters the 
projected volume defined by $1\leq  R/R_{200} \leq 2$, and $|\Delta V| = |\Bar{V}- V_{\rm gal}|\leq 4\,\sigma $,
where $R$ is the galaxy's projected cluster-centric distance, $R_{200}$ is the projected radius within which the mean overdensity is 200 times the critical density of the universe,  $\Bar{V}$ and $V_{\rm gal}$ are the line-of-sight velocity of the cluster and the galaxy's, respectively, and 
$\sigma $ is the radial velocity dispersion.
The $R/R_{200} \leq 2$ limit is due to the size of the field of view of the OmegaWINGS survey.
Values of $\R200 $, $\Bar{V}$ and $\sigma$ are taken from 
\citet{Moretti17}. The general results described in the following sections do not change if galaxies on the 
outskirts of clusters are selected in the range $|\Delta V| \leq 2-8\,\sigma $.

We classify galaxies that meet the criteria of the paragraph above into two categories:
i) we consider to be galaxies infalling alonside filaments (hereafter FG) those that are located
within $45^{\circ}$ from the (projected) direction from the cluster to the other node of 
the filament (hereafter filament direction), and, ii) isotropically infalling galaxies (hereafter IG)  
those that lie outside the filament direction. It should be mentioned that some 
clusters can be nodes of more than one filament. 
For an angle of $45^{\circ}$ and a clustocentric distance of $2R/R_{200}$, the projected average width of the filaments is 3 Mpc, which is the value proposed by \cite{Martinez16} to select the filament's members. We have tested different angles to select filament galaxies and 
found that the general results described in the following sections remain unchanged.

Filaments can vary greatly in shape, extension, width, and whether
they end up in a system or in a galaxy. Filament properties depend on the method used
to identify them, as does the galaxy population that the different methods find inhabiting them 
(see for instance \citealt{Rost:2019}). Our approach probes filaments that are basically straight in 
shape, overdense, and have systems as their ending points.
The number of filaments connected to a cluster scale with cluster mass (e.g. \citealt{Codis:2018,Sarron:2019,Gouin:2019}), 
thus it may be
the case that, in the outskirts of our clusters, we are missing some filaments because: 1) they are 
not dense enough to meet our overdensity criterion, 2) they do not have another system at the other end, or, 3) they do not fit in the cuboid-like regions between systems we use to search for overdensities. 
Therefore, our sample of galaxies in what we call isotropic infall region a:wround our clusters is
expected to be contaminated by galaxies that actually reside in filaments. This must be kept in mind  
for the remaining of the paper, and all differences that we might find between galaxies in filaments and
in the isotropic infall region, are to be considered as lower limits. Therefore, actual differences
would be even more significant.

For comparison purposes, we also consider a sample of galaxies in clusters
consisting in all OW galaxies with projected distance within $\R200 $ of the cluster centre and line-of-sight velocity $|\Delta V|\leq \sigma $. 

We show in Fig. \ref{fig:regions} an example one of the OW clusters in our sample, and
the corresponding galaxies in the three samples described before.

For a complete comparison, we also construct a sample of field galaxies, which includes all 
galaxies in the whole field of the OW clusters with $|\Delta V|>4 \sigma$. 
To avoid a bias in the redshift distribution, we apply a Monte Carlo algorithm that 
randomly selects galaxies with a redshift distribution similar to that of the other 
three samples of galaxies. The number of galaxies by environment and spectral type are 
shown in Table \ref{tab:sample}.

The weighted stellar mass distributions for the total sample and for each spectral type, 
in the four environments considered, are shown in Fig. \ref{fig:mass}.
Median values of the total sample (inner white dots inside the violin plots) are 
$\log(M_{\star}/M_{\odot})=10.02$, $10.18$, $10.06$ and $10.17$ for the field, isotropic infall, filaments and 
clusters, respectively. As can be seen the median values of the stellar mass are similar in all environments. 
Passive galaxies in clusters and filaments have a wider mass range than galaxies in the field and the isotropic 
infall, however the median values across the environments are similar. The only possible exception are 
Post-starburst galaxies that show slightly different medians of stellar mass. It should be noted that this is 
the subsample of spectral types with the least number of galaxies.
For a detailed analysis of the mass distributions and the fractions of galaxies as a function of the spectral type and the cluster-centric distance we refer the reader to \citet{Paccagnella17}.

\section{RESULTS AND DISCUSSION}
\label{sect:results}

 \begin{figure*}
 \begin{center}
\includegraphics[width=15 cm]{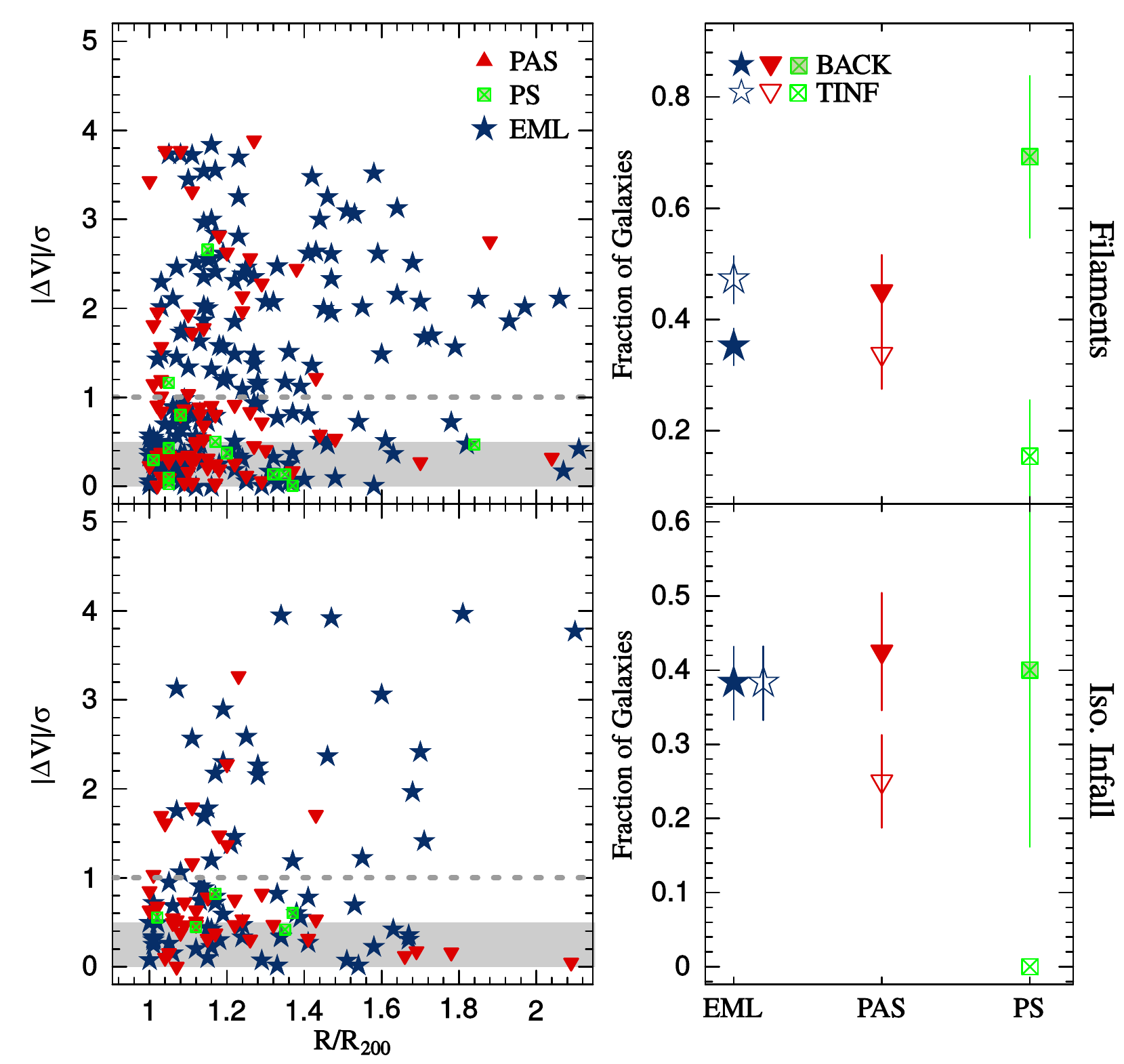}
\caption{\emph{Left} panels show the projected phase-space diagram: the line-of-sight galaxy velocity normalized 
to the line-of-sight cluster velocity dispersion as a function of $R/ \R200 $. Gray regions show the 
backsplash galaxies (BACK). The gray-dashed lines indicate the lower limit of infalling galaxies 
(TINF). Emission-line galaxies are shown as blue stars, passive galaxies in red down triangles, and 
post-starburst in green symbols. \emph{Right} panels show the fraction of galaxies as a function of the 
spectral type. BACK galaxies are shown in solid symbols and TINF in open symbols.
\emph{Top} panels consider galaxies in filaments while \emph{bottom} panels show infalling galaxies.}
 \label{fig:phase}
 \end{center}
\end{figure*}

\subsection{EML and PAS Galaxies}
\label{sect:emlpas}

The fraction of EML and PAS galaxies as a function of the stellar mass 
and environment can be seen in Fig. \ref{fig:frac_mass}. Error-bars in this figure are computed 
with the bootstrap resampling technique (the resampling was performed within each environment). 
We have considered three mass bins: $\log(M_{\star}/M_{\odot})=9.3-10.0$, $10.0-10.7$, $10.7-11.4$. 
The lower and higher mass limits of the first and third mass bins are shown in the bottom panels of Fig. \ref{fig:mass} and delimit the stellar mass completeness. These limits allow a fair
comparison across environments.
Fig. \ref{fig:frac_mass} clearly shows that the environment affects the relative abundance of passive 
and emission line galaxies.  As expected, the fraction of PAS (EML) galaxies increases
(decreases) with mass in all environments, however, environment makes a difference. 
In clusters, the fraction of PAS (EML) galaxies takes the highest (lowest) values, 
followed in decreasing (increasing) order by FG and IG, and the lowest (highest) fraction is seen in the field. 
For the lowest mass bin, the fraction of passive galaxies 
in the field is well under 0.2,  
an indication of the field environment inefficiency to quench these galaxies. 
In the high mass bin, all environments have fractions of 
EML greater than 0.24. 
The fraction of massive PAS galaxies in clusters is larger than 0.7, suggesting that, in addition to the internal processes, the cluster environment is very efficient in quenching these galaxies.

The two environments we explore in the outskirts of clusters show fractions of PAS and EML with 
values between those of the field and clusters. The comparison between IG and FG shows that, for the two lowest mass bins, the PAS (EML) fraction is higher (lower) for galaxies that are infalling 
along filaments than those that are being accreted isotropically. These results are in agreement with 
the findings of \citet{Martinez16} and \citet{Salerno2019} on the outskirts of groups of galaxies 
at low and high redshift, respectively. 
We use the test proposed by \citet{Muriel:2014} to evaluate whether the observed differences between the fractions determined for FG and IG are statistically significant. This test computes the bin-to-bin 
cumulative differences between the two samples along the stellar mass range, and checks whether the 
resulting quantity is consistent with the null hypothesis of the two samples drawn from the same 
population. This test was only applied to the comparison between galaxies in the isotropic infall and in 
filaments, since in the other two environments (field and clusters) the differences in the fractions are 
much larger than the quoted errors. 
Results of the test are shown in the upper right part of both panels in Fig. \ref{fig:frac_mass}, in terms 
of the rejection probability of null hypothesis ($R_{\rm p}$). In both cases, the null hypothesis is 
overruled. i.e., galaxies infalling in the filament direction are statistically more quenched that those 
infalling isotropically. If the above analysis is repeated using a mass limited sample ($\log(M_{\star}/M_{\odot})=9.8$, \citealt{Vulcani11}) the results do not change, although they become noisier. Since our sample of galaxies in the outskirts of clusters was selected prioritizing completeness over purity, the reported differences between galaxies in the direction of filaments and in the isotropic infall, should be taken as a lower limit.

Our results indicate that the outskirts of clusters can pre-process galaxies, contributing significantly 
to the suppression of their star formation, and also that filaments are more effective in doing so than 
the isotropic infall region. 
Fig. \ref{fig:frac_mass} can provide a raw estimation of how effective to quench galaxies 
these two environments are.
Let us consider the abundance of PAS galaxies as a function of 
the stellar mass ({\em left panel} of Fig. \ref{fig:frac_mass}). 
Furthermore, let us also assume the
field is inefficient to environmentally quench galaxies. 
Thus, mass quenching is the only responsible
of the existence of PAS galaxies in the field, which means that up to a fraction 
$F_{\rm PAS}^f(M_i)= 0.14,~ 0.25,$ and $0.32$ of the galaxies in the field 
(blue squares in the {\em left panel} of Fig. \ref{fig:frac_mass}) have been mass 
quenched in our three mass bins ($i=1,2,3$), respectively.
This is a strong assumption since our field sample may include small groups 
of galaxies which, in turn, would introduce an environmental quenching factor that, 
however small, we can not quantify. Therefore these fractions should be better 
considered as upper limits to the mass quenching efficiency.
These numbers reflect the mass quenching efficiency to the present time. 
The `reservoir' of galaxies that have not been quenched by mass so far, 
and are therefore available to be quenched by the environment is measured by the 
fraction of EML galaxies in the field: 
$F_{\rm EML}^f(M_i)= 0.86,~0.75,$ and $0.68$, respectively.
Let us consider now an environment $x$, which
is denser than a reference environment. In our case we take the field as reference 
environment due to its inefficiency to environmentally quench galaxies. 
For galaxies of mass $M_i$, the quenching 
efficiency of the environment $x$, relative to the field, is measured as 
\citep{Peng10}:
\begin{equation}
    Q^x(M_i)=\frac{F_{\rm PAS}^x(M_i)-F_{\rm PAS}^f(M_i)}{F_{\rm EML}^f(M_i)}.
\end{equation}
This ratio is the fraction of galaxies of mass $M_i$ that have been 
quenched in the environment $x$, but would be star forming if they were inhabiting 
the field instead.
The fraction of PAS galaxies in the isotropic infall region (filaments) is
$F_{\rm PAS}^{\rm{iso}}(M_i)= 0.25,~0.36$, and $0.57$ 
($F_{\rm PAS}^{\rm{fil}}(M_i)= 0.33,~0.48$ and $0.56$), 
and represent the fraction of galaxies in the mass bins $i=1,2,3$ that, at the present time, are quenched prior 
to their eventual entrance to the clusters. 
These numbers yield relative quenching efficiency for the isotropic infall region is
$Q^{\rm{iso}}(M_i)= 0.11, ~0.15$, and $0.37$. Likewise, the filaments have a
relative quenching efficiency of $Q^{\rm{fil}}(M_i)= 0.22, ~0.31$ and $0.35$. 
Note that PS have not been taken into account in the computation of these fractions. Since they are already turning passive, they could have been considered as such.
They are negligible in the field sample, and account only for 0.04, and 0.06, of the samples of isotropic infall and filament galaxies respectively, over the whole 
mass range. Their inclusion as passive galaxies would not change our rough numbers
in a significant way.

These results suggest that high mass galaxies in the 
outskirts of clusters are more likely to be affected by both, mass and environmental quenching, than low 
mass galaxies.  It is also interesting to note that, for the three mass bins, the fraction of galaxies 
affected by environmental quenching in the infall region is similar to the mass quenching, while it is 
higher in filaments (at least for the first two bins).

\subsection{Post-starburst galaxies}
\label{sect:ps}

A powerful way to study the influence of the environment on the evolution of galaxies is to study the 
presence of galaxies that are in transition from star-forming to quiescent. From the point of view of 
the spectroscopy, this can be done by studying the presence of E + A galaxies (\citealt{Dressler:83}) that are known as post-starburst galaxies (PS). \cite{Fritz2014} use the spectral types k+a 
and a+k to  classify PS galaxies. These are galaxies that are supposed to have stopped the star formation 
some $10^{7}$ years ago. Among the cluster members, \citet{Paccagnella17} find that PSs in OmegaWINGS are a combination of galaxies with a mix of times since infall, including both, backsplash and virialized galaxies. They also find that the frequency of PS galaxies in clusters is much higher than in the field. Similar results are found by \citet{Paccagnella19} who report a frequency of PS galaxies  growing from isolated galaxies, to binary systems, groups and clusters, with a progressive increase in their fraction with the halo mass.

As can be seen in Table \ref{tab:sample}, the fraction of PS galaxies increases as we 
move from the field to denser environments.
The percentage of PS in the isotropic infall is slightly smaller than in the filament region. However, 
the small number of PS galaxies in our samples of field and isotropic infall region prevent us from 
having a reliable estimation of the uncertainty involved, which does not allow us to assess the significance
of the result.
Our data indicate that the fraction of PS in filaments is smaller than in clusters. 

Several authors have suggested that the ram-pressure stripping is the most effective mechanism to create PS 
galaxies in clusters (\citealt{Dressler:83}; \citealt{Couch1987}; \citealt{Dressler92}; \citealt{Poggianti99}; \citealt{Balogh:2000}; \citealt{Tran03};  \citealt{tran04}; 
\citealt{Tran07}; \citealt{Poggianti09}; \citealt{fritz14}; \citealt{Paccagnella17}; \citealt{Paccagnella19}). The star formation that was previously triggered during the infall into the cluster, is then 
abruptly extincted when the galaxy enters the innermost regions of the cluster.
If this were the case, PS galaxies that are located on the outskirts of clusters should have previously 
orbited close to the cluster centre, becoming what are known as backsplash galaxies. This hypothesis may be valid if some galaxies can cross the cluster and go back to the outskirts in a time lesser than or equal to the duration of the PS stage (1-1.5 Gyrs). \citet{Martinez:2020} use numerical simulations to analyse the galaxy orbits in clusters. The authors find that, on average, backsplash galaxies take 1.6 Gyrs since crossing the cluster's virial radius on their way in until they cross it again on their way out. These results suggest that almost half of the galaxies that have been transformed into PS during the infall and have subsequently left the cluster as backsplash galaxies, would still show characteristics of PS galaxies.
In order to test the 
hypothesis that some of the PS galaxies in outskirts of clusters are backsplash, we apply the criterion proposed by \citet{Muriel:2014} to select both, backsplash and infaller 
galaxies. For these authors, galaxies in the outskirts of clusters ($R\geq R_{200}$) with 
$|\Delta V|/\sigma \leq 0.5$ have a high probability of being backsplash (hereafter BACK). 
On the contrary, galaxies in the outskirts with  
$|\Delta V|/\sigma > 1$ are considered as true infalling galaxies (hereafter TINF). Note that according to our selection in velocity, not all the galaxies in the outskirts are classified as BACK or TINF. The \emph{left} panel of 
Fig. \ref{fig:phase} shows the phase-space diagram for the galaxies in our samples. In Table
\ref{tab:phasedata} and in the {\em right} panel of Fig. \ref{fig:phase} we present the fraction of BACK and 
TINF 
galaxies as a function of the spectral classification and environment in the outskirts of clusters. 
As can be seen in Table
\ref{tab:phasedata},  the majority (69\%) of PS galaxies in the filament region are BACK. Only 15\% of the 
PS galaxies are TINF. For the same environment,  35\% of the EML galaxies are BACK while 47\% are TINF. PAS 
galaxies show intermediate behaviour, with 45\% of BACK and 33\% TINF.  These numbers suggest that an important fraction of the PS galaxies in the filament region in the outskirts of clusters were 
previously at smaller clustocentric distances. In their raid through the cluster, galaxies could have been quenched by ram pressure stripping or any of the processes described in section \ref{sect:Intro}. In the isotropic infall region the results show a 
similar tendency, however, the low number of PS galaxies in this region makes it very difficult to get conclusive results.

If most of the PS galaxies in the filament region are actually BACK, 
they are more likely to have been quenched in the clusters, regardless of their original infall
direction. This could suggest that the filament region (and probably also the infall region) has been inefficient to produce a rapid quenching to star forming galaxies 
during the last $\sim1.5$ Gyr. Therefore, many of the passive galaxies that we observe falling to clusters, may have been quenched by mechanisms that act over
longer time scales, or these galaxies were quenched a long time before.
Quenching in filaments as early as $z\sim0.9$ has been reported by \citet{Salerno2019}. 
They present evidence that a significant fraction of the galaxies in filaments have already been
pre-processed by that redshift.

\begin{table}
\centering
\begin{tabular}{l|l|c|c|c}
\hline
           &                 &  PAS             &   EML           &  PS \\
\hline           
Filaments   & BACK      &  $0.45\pm 0.04$ & $0.35\pm 0.06$  & $0.69\pm 0.12$  \\
            & TINF      &  $0.33\pm 0.04$ & $0.47\pm 0.06$  & $0.15\pm 0.09$  \\ 
            \hline
Iso. Infall & BACK      & $0.42\pm 0.06$ & $0.38\pm 0.08$   & $0.40\pm 0.20$    \\
            & TINF      & $0.25\pm 0.05$ & $0.37\pm 0.07$   & $0.00\pm 0.00$     \\
\hline               
\end{tabular}
\caption{Fraction of BACK and TINF galaxies as a function of the spectral type and environment, 
see Fig. \ref{fig:phase}.}
\label{tab:phasedata}
\end{table}


\section{Summary}
\label{sect:conclu}

In this work we study the effects of environment on galaxy quenching at $0.04 < z < 0.08$ 
using data from WINGS and OmegaWINGS cluster surveys. We focus our study in the outskirts of galaxy 
clusters, i.e. projected distances $1\le R/R_{200}\le 2$, and distinguish between two regions according to the clustocentric 
direction: filaments and the isotropic infall. 

Firstly, we search for filamentary structures linking galaxy groups/clusters identified by
\citet{Lim17} over the 6dFGRS DR3 \citep{Jones09}. We follow \citet{Martinez16} in the filament 
identification process. We then cross-match groups/clusters that are nodes of filaments,
with OW clusters. This matching results in 14 OW clusters that are linked to other groups/clusters
by filaments. 

We study the fraction of passive and emission-line galaxies in four different environments: 
clusters, filaments, the isotropic infall region of clusters and the field.
We find a fraction of passive galaxies in the outskirts of clusters intermediate between that
of the clusters and the field's. Furthermore, we find evidence of a more effective quenching in the
direction of the filaments.
These results
are in agreement with an scenario in which pre-processing of galaxies prior to their entrance
to clusters is more effective in filaments \citep{Martinez16,Salerno2019}.

We also analyse the abundance of post-starburst galaxies in the outskirts of clusters.
Since post-starburst galaxies are believed to be the result of strong ram-pressure stripping
in galaxies, and this effect is strongest in the innermost regions of clusters, we use
the phase-space position of galaxies in the outskirts of clusters to focus our analysis in
two sets of galaxies: backsplash and true infallers. This classification allows us to pick up
galaxies that are likely to have passed by the inner regions of clusters and compare them
with galaxies that are likely to never have been within one virial radii of the cluster centre.
We find that up to $~70\%$ of post-starburst galaxies in the direction of filaments are likely
to be backsplash, this number drops to $\sim40\%$ in the isotropic infall region.

The low fraction of galaxies in filaments that have been recently quenched (post starburst galaxies) 
and are falling into clusters for the first time, support the scenario in which an 
important fraction of filament galaxies have been quenched long time ago.

\section*{Acknowledgements}

This paper has been partially supported with grants from Consejo Nacional de 
Investigaciones Cient\'ificas y T\'ecnicas (PIP 11220130100365CO) Argentina, 
and Secretar\'ia de Ciencia y Tecnolog\'ia, Universidad Nacional de C\'ordoba, Argentina.




\bibliographystyle{mnras}
\bibliography{mnras} 


\appendix


\bsp	
\label{lastpage}
	\end{document}